\documentclass{article}
\usepackage{spconf,amsmath,graphicx,url,times,booktabs,tabularx}
\usepackage{bm}
\usepackage{amssymb}
\usepackage{cite}
\usepackage{url}
\usepackage{setspace}

\usepackage{multirow}

\DeclareMathOperator{\encoder}{Enc}
\DeclareMathOperator{\decoder}{Dec}
\DeclareMathOperator{\vggish}{VGGish}
\DeclareMathOperator{\embed}{Embed}
\DeclareMathOperator{\gpt}{GPT-2}
\DeclareMathOperator{\mh}{MultiHeadAttention}
\DeclareMathOperator{\lmhead}{LMHead}

\DeclareMathOperator{\lin}{Linear}

\title{Audio Captioning using Pre-Trained Large-Scale Language Model\\Guided by Audio-based Similar Caption Retrieval}  

\name{Yuma Koizumi, Yasunori Ohishi, Daisuke Niizumi, Daiki Takeuchi, and Masahiro Yasuda}
\address{NTT Corporation, Tokyo, Japan}

\def\Hline{
\noalign{\ifnum0=`}\fi\hrule \@height 4.\arrayrulewidth \futurelet
\reserved@a\@xhline}

\begin{document}
\ninept
\maketitle

\begin{abstract}
\vspace{-4pt}
The goal of audio captioning is to translate input audio into its description using natural language. One of the problems in audio captioning is the lack of training data due to the difficulty in collecting audio-caption pairs by crawling the web. In this study, to overcome this problem, we propose to use a pre-trained large-scale language model. Since an audio input cannot be directly inputted into such a language model, we utilize guidance captions retrieved from a training dataset based on similarities that may exist in different audio. Then, the caption of the audio input is generated by using a pre-trained language model while referring to the guidance captions. Experimental results show that (i) the proposed method has succeeded to use a pre-trained language model for audio captioning, and (ii) the oracle performance of the pre-trained model-based caption generator was clearly better than that of the conventional method trained from scratch.
\end{abstract}


\vspace{-4pt}
\section{Introduction}
\label{sec:intro}
\vspace{-2pt}

Audio captioning is a crossmodal translation task when translating input audio into its description using natural language~\cite{ac1,ac2,ac3,audiocaps,clotho,ntt_task6,Interspeech2020Koizumi,DCASE2020_Takeuchi}.
This task potentially enhances the level of sound environment understanding from merely tagging events~\cite{aed,aed2} (e.g.\,alarm), scenes~\cite{asc} (e.g.\,kitchen) and conditions~\cite{asd} (e.g.\,normal/anomaly) to higher contextual information including concepts and/or high-level knowledge.
For instance, a smart speaker with an audio captioning system will be able to output ``{\it a digital alarm in the kitchen has gone off three times},'' and might be the first step for giving us more intelligent recommendations such as ``{\it turn the gas range off}.''

One of the challenges in audio captioning is to address the lack of training data~\cite{DCASE2020_Takeuchi}.
Typical datasets in audio captioning, AudioCaps~\cite{audiocaps} and Clotho~\cite{clotho}, contain only 49,838 and 14,465 training captions, respectively, whereas there are 36M training sentence-pairs in the WMT 2014 English-French dataset for machine translation. It is due to the difficulty in collecting audio and the corresponding captions by crawling the web.

Generally, to overcome such a problem, task-specific fine-tuning of a pre-trained model is a promising strategy. In audio event detection and scene classification, several pre-trained models, such as VGGish \cite{vggish}, $L^3$-Net \cite{openl3}, and COALA~\cite{coala}, have been published to achieve better results with less training data.
Similarly, in NLP, large-scale pre-trained language models, such as the Bidirectional Encoder Representations from Transformers (BERT) \cite{bert} and Generative Pre-trained Transformer (GPT) \cite{gpt,gpt2}, also improve the performance of various tasks.

Unfortunately, it is difficult to straightforwardly adopt such powerful pre-trained language models for audio captioning. This is because such models assume that the input is also a word sequence, not an audio sequence.
To benefit from such pre-trained language models in crossmodal translation tasks such as audio captioning, we state a research question as "{\it how can another modal information be inputted to a pre-trained language model?}"

\begin{figure}[t]
  \centering
\includegraphics[width=85mm,clip]{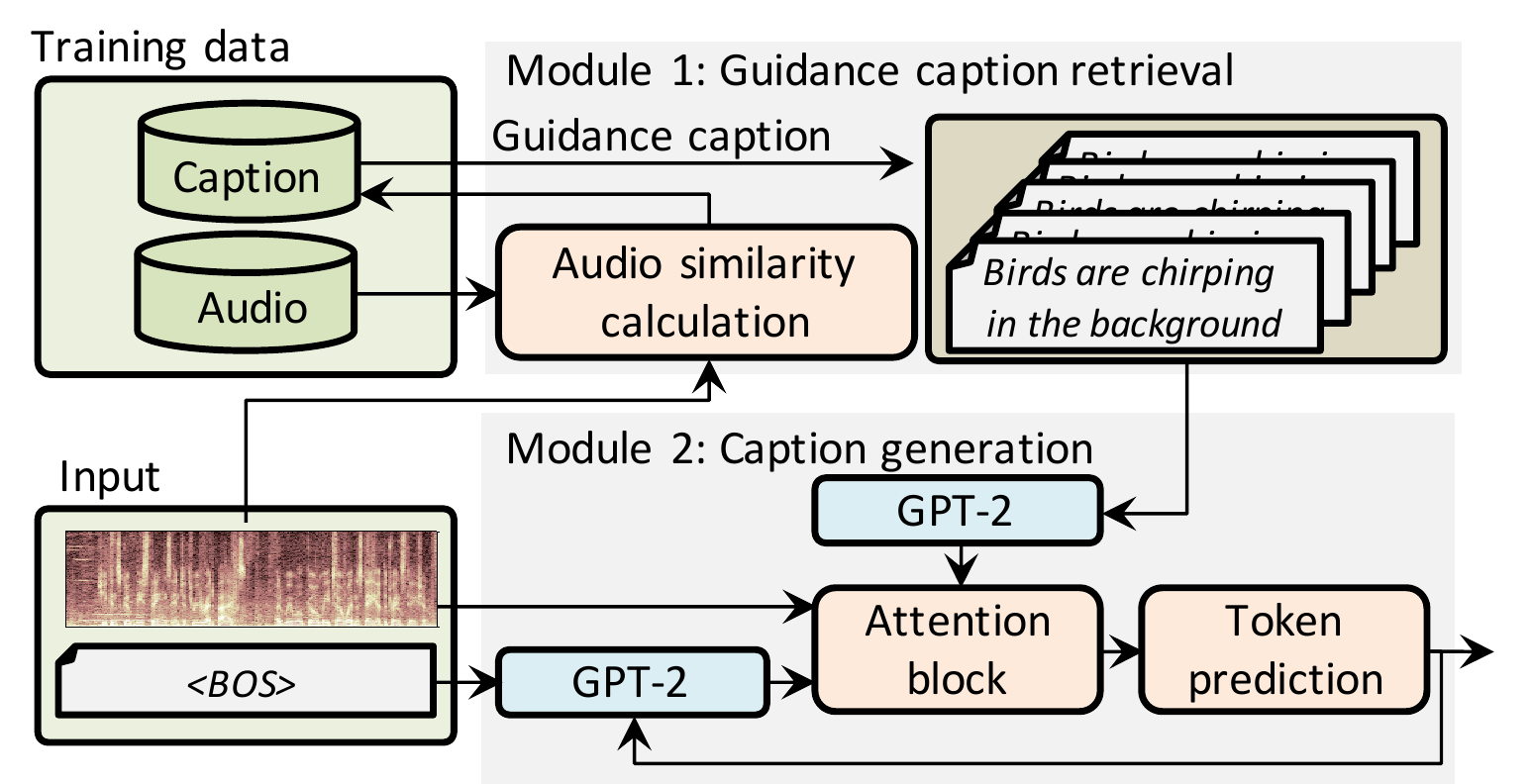} 
  \vspace{-5pt}
  \caption{Overview of the proposed audio captioning system. The orange blocks have trainable blocks and the blue blocks denote frozen pre-trained models.}
  \label{fig:ov}
  \vspace{-5pt}
\end{figure}

This paper is the first study to utilize a large-scale pre-trained language model for the audio caption task. 
The proposed method is a cascaded system consisting of two modules as shown in Fig.\,\ref{fig:ov}.
The first module works like an encoder; this module outputs guidance captions for the next module in which these captions are retrieved from the training dataset based on the similarity between the input audio and training samples. The second module works like a decoder; this module generates the caption of the input audio using a pre-trained language model while referring to the retrieved captions.
This strategy enables us to avoid directly inputting the audio to the pre-trained language model, and to utilize it for audio captioning.

\vspace{-4pt}
\section{Preliminaries of audio captioning}
\label{sec:conv}
\vspace{-4pt}

Audio captioning is a task to translate an input audio sequence $\bm{\Phi} = ( \bm{\phi}_1, ..., \bm{\phi}_T)$ into a (sub-)word token sequence $(w_1,...,w_N)$. Here, $\bm{\phi}_t \in \mathbb{R}^{D_a}$ is a set of acoustic features at time index $t$, and $T$ is the length of the input sequence. 
Each element of the output $w_n \in \mathbb{N}$ denotes the $n$-th token's index in the word vocabulary, and $N$ is the length of the output sequence.

Previous studies addressed audio captioning through the Encoder-Decoder framework such as sequence-to-sequence models (seq2seq) \cite{seq2seq1,seq2seq2,DCASE2020_Takeuchi} and/or Transformer \cite{transformer,Interspeech2020Koizumi}.
First, the encoder embeds $\bm{\Phi}$ into a feature-space as $\bm{\nu}$. 
Then, the decoder predicts the posterior probability of the $n$-th token while using conditioning information $\bm{\nu}$ and 1st to $(n-1)$-th outputs recursively as
\begin{align}
\bm{\nu} 
&= \encoder \left( \bm{\Phi} \right),
\label{eq:encoder}\\
p( w_n | \bm{\Phi}, \bm{w}_{n-1} )
&= \decoder \left( \bm{\nu}, \bm{w}_{n-1} \right),
\label{eq:decoder}
\end{align}
where $\bm{w}_{n-1} = (w_1,...,w_{n-1})$, and $w_n$ is estimated from the posterior.
To improve the word prediction accuracy, some studies estimate additional information such as keywords~\cite{audiocaps,Interspeech2020Koizumi}, and pass it to the decoder as conditioning information.
Then, (\ref{eq:decoder}) becomes 
$p( w_n | \bm{\Phi}, \bm{w}_{n-1} )
= \decoder_{\theta_d} \left( \bm{\nu}, \bm{c}, \bm{w}_{n-1} \right)$
where $\bm{c}$ is the additional condioning information.

The decoder is an autoregressive language model for generating a caption while conditioned by $\bm{\Phi}$. It is known that the sentence generation accuracy can be dramatically improved by using an autoregressive pre-trained language model such as GPT~\cite{gpt,gpt2}. However, previous audio captioning studies did not use them and trained the decoder from scratch~\cite{ac1,ac2,ac3,audiocaps,clotho,ntt_task6,Interspeech2020Koizumi,DCASE2020_Takeuchi}. This is because such a pre-trained language model estimates $p( w_n | \bm{w}_{n-1} )$, that is, it is not able to condition the posterior by using other modal's information such as $\bm{\nu}$.

\vspace{-4pt}
\section{Proposed Method}

\vspace{-4pt}
\subsection{Basic strategy}

We construct an encoder to convert an input audio sequence to a token sequence, and use it for conditioning a pre-trained language model.
That is, in contrast to the conventional audio caption framework such as (\ref{eq:encoder}) and (\ref{eq:decoder}), 
the encoder outputs a reference token sequence
$\bm{w}^{\mbox{\scriptsize ref}} = 
(w^{\mbox{\scriptsize ref}}_1, ..., w^{\mbox{\scriptsize ref}}_M)$
and 
the decoder predicts the posterior while using $\bm{w}^{\mbox{\scriptsize ref}}$ as conditioning information as
\begin{align}
\bm{w}^{\mbox{\scriptsize ref}}
&= \encoder \left( \bm{\Phi} \right),
\label{eq:encoder_p}\\
p( w_n | \bm{\Phi}, \bm{w}_{n-1} )
&= \decoder \left( \bm{w}_{n-1}, \bm{w}^{\mbox{\scriptsize ref}} \right),
\label{eq:decoder_p}
\end{align}
where $M$ is the length of $\bm{w}^{\mbox{\scriptsize ref}}$, and a part of the decoder consists of a pre-trained language model.

Here, the encoder does not necessarily need to ``generate'' a token sequence, rather only needs to output token sentences that are useful for the decoder conditioning.
Based on this idea, we construct the encoder to retrieve appropriated (similar) captions from given training dataset based on audio similarity as shown in Fig.\,\ref{fig:ov}. Hereafter, we call such retrieved captions as ``guidance captions''.
In the following sections, we describe the ``encoder'' step in Sec.\,\ref{sec:cap_ret} and the ``decoder'' step in Sec.\,\ref{sec:decoder}.

\vspace{-4pt}
\subsection{Audio-based guidance caption retrieval}
\label{sec:cap_ret}

\begin{figure}[t]
  \centering
\includegraphics[width=\linewidth,clip]{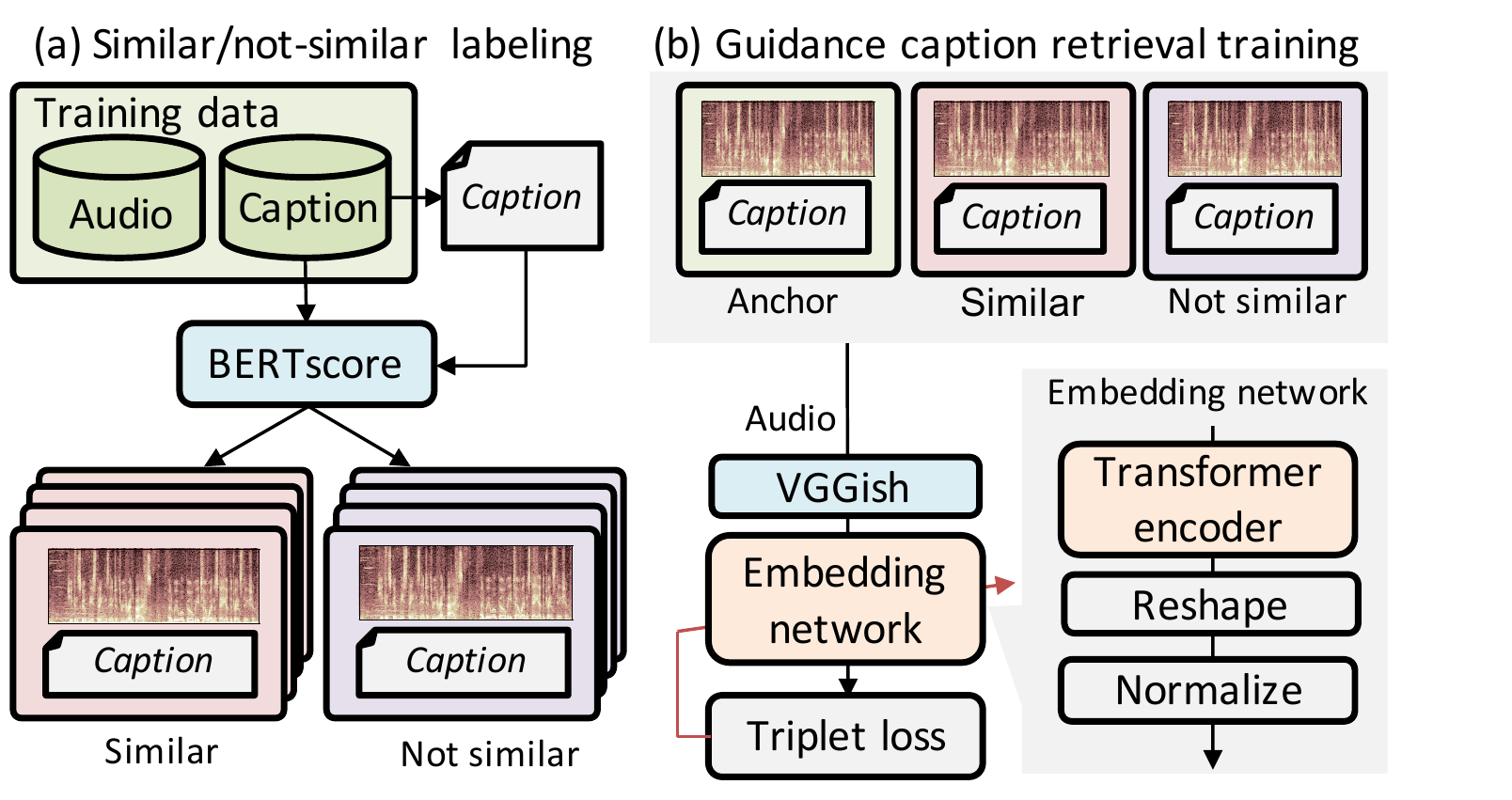} 
  \vspace{-15pt}
  \caption{Training procedure of audio-based guidance caption retrieval.}
  \label{fig:triplet}
  \vspace{-0pt}
\end{figure}

The goal of the ``encoder'' step is to retrieve guidance captions from the training dataset based on audio similarity.
This audio similarity should take high value when the two captions are similar even if its corresponding audios are not similar.
In order to achieve this requirement, the training of this step consists of (i) defining the sentence similarity between captions in the training dataset, and (ii) training a DNN to predict the similarity from the audio.

As the similarity between two captions, we use BERTScore~\cite{BERTScore} as shown in Fig.\,\ref{fig:triplet} (a), because the BERTScore correlates good with human judgments in evaluation for image captioning.
The detailed procedure is followings:
\begin{enumerate}
  \setlength{\parskip}{0pt}
  \setlength{\itemsep}{0pt}
    \item Calculating the BERTScore between all the possible caption pairs in the training dataset.
    \item Normalizing the BERTScore so that the maximum and minimum values are 1 and 0, respectively.
    \item Labeling captions whose normalized BERTScore is larger than a pre-defined threshold as ``similar'', and ``not similar'' for others. We used 0.7 as the threshold.
\end{enumerate}

The procedure of guidance caption retrieval is followings.
First, the audio input in the time-domain $\bm{x}$ is converted into the embedded space using a DNN, and then the distances between the input and all audio samples in the training dataset are calculated.
The DNN consists of freezed VGGish~\cite{vggish} and a trainable embedding network as shown in Fig.\,\ref{fig:triplet} (b). First, $\bm{x}$ is converted to its feature sequence $\bm{\Phi}$ as
\begin{align}
\bm{\Phi} = \vggish( \bm{x} ),
\end{align}
where 
$\bm{\Phi} \in \mathbb{R}^{D_a \times T}$ and $D_a=128$ in VGGish.
Then, $\bm{\Phi}$ is converted to a fixed dimension vector $\bm{e}$ as
\begin{align}
\bm{e} = \embed( \bm{\Phi} ).
\end{align}
The trainable layer of $\embed(\cdot)$ is only one Transformer-encoder layer which consists of a multi-head self-attention layer and a feed-forward block.
After the Transformer-encoder layer, we use a reshape function $\mathbb{R}^{D_a \times T} \to \mathbb{R}^{D_aT}$. Finally, we normalize $\bm{e}$ so that $\lvert \bm{e} \rvert = 1$. In this study, we assume the length of the time-domain input is always the same, thus the dimension of $D_aT$ of the input and all training samples is also always the same.
To retrieve guidance captions, we use $\ell_2$ distances $\mathcal{D} (\bm{a}, \bm{b}) = \lVert \bm{a} - \bm{b} \rVert_2^2$ between the embedded features of the input audio and all the audio in the training dataset.
Then, this module outputs the top $K=5$ captions with smaller distances as $\bm{w}^{\mbox{\scriptsize ref}}$.

To train the embedding network, we adopted the triplet-based training~\cite{triplet} according to the success of this strategy in crossmodal retrieval~\cite{linguistic6,YasudaInterspeech}.
The network is trained to minimize the triplet loss~\cite{triplet} as
\begin{align}
\mathcal{L} = 
\max \left( 0,
\mathcal{D}\left( \bm{e}^{a}, \bm{e}^{p} \right) -
\mathcal{D}\left( \bm{e}^{a}, \bm{e}^{n} \right) +
\alpha
\right),
\end{align}
where $\alpha = 0.3$ is a margin parameter,
$\bm{e}^{a}$, $\bm{e}^{p}$, and $\bm{e}^{n}$ are the embeddings of
the anchor, positive, and negative samples, respectively.
In this study, 
the anchor is the input audio and
the positive is a randomly selected audio in training data which labeled as ``similar'' to the anchor in accordance with the above BERTScore-based similarity.
As the negative sample, we selected a semi-hard negative sample~\cite{semihard1} from ``not similar'' samples which satisfy with
\begin{align}
\mathcal{D}\left( \bm{e}^{a}, \bm{e}^{p} \right) \leq
\mathcal{D}\left( \bm{e}^{a}, \bm{e}^{n} \right) <
\mathcal{D}\left( \bm{e}^{a}, \bm{e}^{p} \right) + \alpha.
\end{align}

\vspace{-4pt}
\subsection{Caption generation using GPT-2 decoder}
\label{sec:decoder}

\begin{figure}[t]
  \centering
\includegraphics[width=\linewidth,clip]{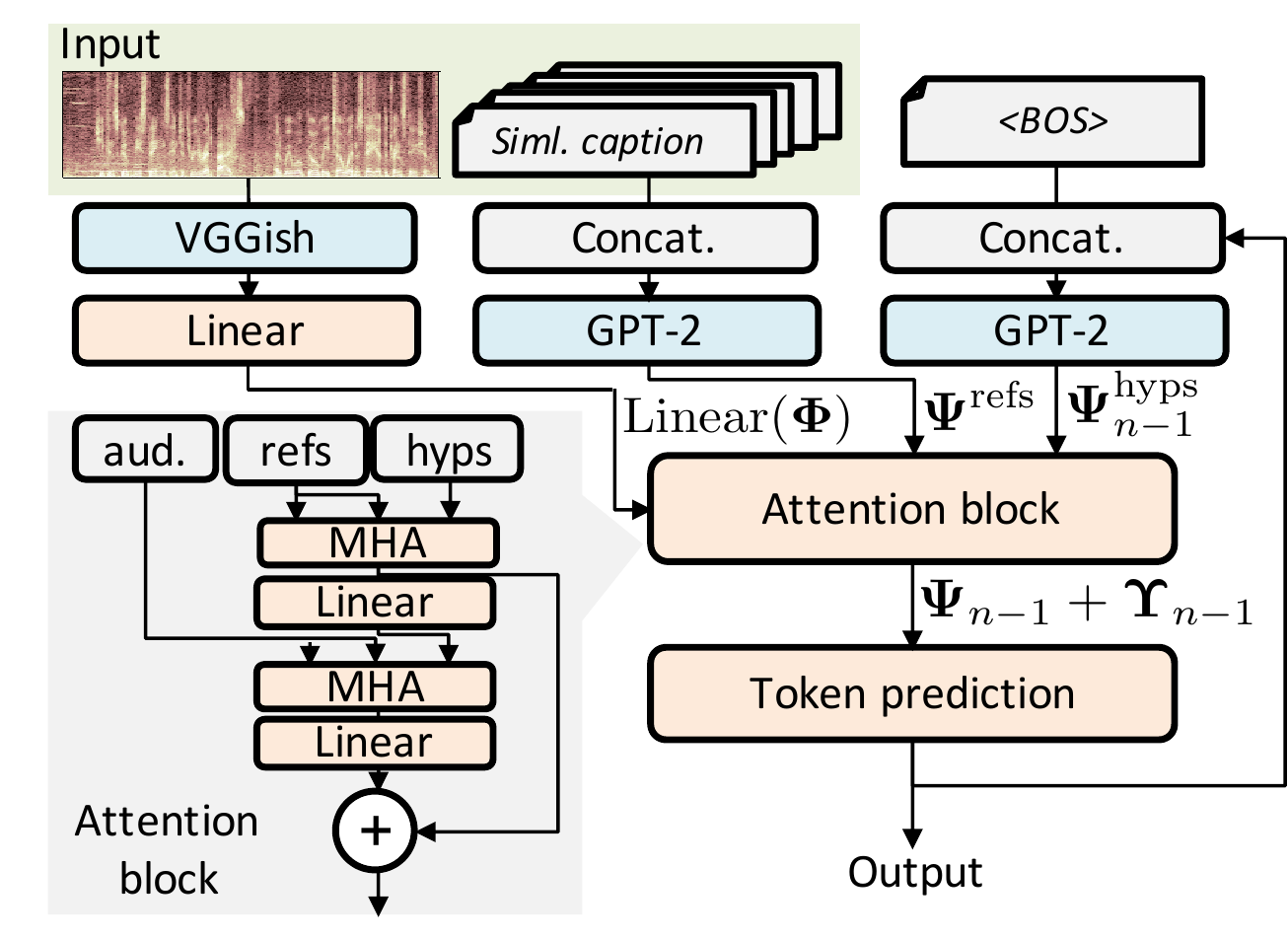} 
  \vspace{-10pt}
  \caption{Flowchart of caption generation using GPT-2. MHA denotes multi-head attention layer.}
  \label{fig:decoder}
  \vspace{-5pt}
\end{figure}

As the pre-trained model in the decoder, we use GPT-2~\cite{gpt2} because of its powerful sentence generation capability.
In addition, GPT-2 is easy to use as a part of the decoder because it is an autoregressive language model. 
The most intuitive and simplest way is that GPT-2 predicts the next sentence of concatenated $K$ guidance captions.
However, this strategy does not work well.
The reason is that since guidance captions are independent $K$ sentences, the concatenated guidance captions will seem to be a {\it strange} document.
Such strange documents might confuse GPT-2 that trained using {\it natural} documents, and be a cause to generate strange caption.

Figure\,\ref{fig:decoder} shows the flowchart of caption generation.
We use GPT-2 without the final token prediction layer as a feature extractor.
That is, the concatenated guidance captions and target captions passed to the frozen GPT-2 independently. Then, we integrate them using trainable multi-head attention layers.
In more detail, 
we first pass 
$\bm{w}^{\mbox{\scriptsize ref}}$
and
$\bm{w}_{n-1}$ to GPT-2 independently as
\begin{align}
\bm{\Psi}^{\mbox{\scriptsize refs}} &= \gpt \left( \bm{w}^{\mbox{\scriptsize ref}} \right),\\
\bm{\Psi}^{\mbox{\scriptsize hyps}}_{n-1} &= \gpt \left( \bm{w}_{n-1} \right),
\end{align}
where
$\bm{\Psi}^{\mbox{\scriptsize refs}} \in \mathbb{R}^{D_l \times M}$ and 
$\bm{\Psi}^{\mbox{\scriptsize hyps}}_{n-1} \in \mathbb{R}^{D_l \times (n-1)}$, here $D_l = 768$ in the case of the standard GPT-2 (a.k.a.\,117M). 
Then, these matrices are integrated by using a multi-head attention layer as
\begin{align}
\bm{\Psi}_{n-1} = \mh \left( \bm{\Psi}^{\mbox{\scriptsize hyps}}_{n-1}, \bm{\Psi}^{\mbox{\scriptsize refs}} \right),
\end{align}
where $\mh (\bm{a}, \bm{b})$ uses $\bm{a}$ as {\it query} and $\bm{b}$ as {\it key} and {\it value}, and 
$\bm{\Psi}_{n-1} \in \mathbb{R}^{D_l \times (n-1)}$.
In addition, we also integrate the audio-based feature into $\bm{\Psi}_{n-1}$ using the a multi-head attention layer.
In order to reduce the number of parameters, the dimension of both $\bm{\Psi}_{n-1}$ and $\bm{\Phi}$ are reduced to $D_r = 60$ using Linear layers
as $\bm{\Psi}_{n-1}'$ and $\bm{\Phi}'$, then passed them to the multi-head attention layer as
\begin{align}
\bm{\Upsilon}_{n-1} 
= \lin\left( 
\mh \left( \bm{\Psi}_{n-1}', \bm{\Phi} '\right) \right),
\end{align}
where $\lin(\cdot)$ adjusts the dimension to $\mathbb{R}^{D_l}$.
Finally, the integrated feature is passed to the pre-trained token prediction layer $\lmhead(\cdot)$ to obtain the posterior as
\begin{align}
p( w_n | \bm{\Phi}, \bm{w}_{n-1} ) = \lmhead\left( \bm{\Psi}_{n-1} + \bm{\Upsilon}_{n-1} \right).
\end{align}
Note that $\lmhead(\cdot)$ is also trainable to fit the decoder for statistics of the words available in training captions.

In the training phase, 
we used randomly selected and ordered $K=5$ captions from ``similar'' labeled captions in the training dataset as $\bm{w}^{\mbox{\scriptsize ref}}$.
The loss function was the smoothed cross-entropy with teacher forcing, that is, we calculated posterior $p( w_n | \bm{\Phi}, \bm{w}_{n-1} )$ using the ground-truth $\bm{w}_{n-1}$ and minimize the cross-entropy between the true $n$-th word $w_n$ and the posterior with label smoothing parameter $\lambda = 0.1$.

\vspace{-4pt}
\section{Experiments}
\label{sec:exp}

\vspace{-6pt}
\begin{table*}[ttt]
\caption{Experimental results on AudioCaps dataset}
\label{tab:example}
\centering
\begin{tabular}{ c| l | ccccccccc }
\toprule
\textbf{Scope} 	& \textbf{Method} 	& \textbf{B-1}	& \textbf{B-2}	& \textbf{B-3}	& \textbf{B-4}	& \textbf{METEOR}	& \textbf{CIDEr}	&  \textbf{ROUGE-L} & \textbf{SPICE} \\	
\midrule
\multirow{2}{*}{(i)} &
{\tt TopDown-AlignedAtt(1NN)}~\cite{audiocaps}	    
& 61.4		& 44.6		& 31.7			        & {\bf 21.9}	& {\bf 20.3}    & {\bf 59.3}	& {\bf 45.0}	& {\bf 14.4} \\
 & {\tt Ours}	        
& {\bf 63.8}	& {\bf 45.8}	& {\bf 31.8}	& 20.4			& 19.9			& 50.3			& 43.4	        & 13.9 \\
\midrule 
\multirow{3}{*}{(ii)} & {\tt 1NN-VGGish}	\cite{audiocaps} 
& 44.2		& 26.5		& 15.8			& 9.0			& 15.1			& 25.2			& 31.2	& 9.2 \\
 & {\tt Triplet-Retrieval}	 
& {\bf 47.5}		& {\bf 28.8	}	& {\bf 17.7}			& {\bf 10.4}			& {\bf 16.2}			& {\bf 30.6}			& {\bf 33.1}	& {\bf 10.9} \\
& {\tt Top1-BERTScore}	 & 74.0		& 59.9		& 48.2			& 38.1			& 29.1			& 95.2			& 59.2	& 18.6 \\
\midrule 
\multirow{3}{*}{(iii)} 
 & {\tt AlignedAtt(GT)-VGGish-LSTM} \cite{audiocaps}	
& 69.1		& 52.3		& 36.4			& 26.1			& 23.6			& 77.7			& 49.6	& 17.2  \\
 & {\tt Ours (GT)}	
& {\bf 75.1}	& {\bf 60.4}		& {\bf 48.0}			& {\bf 37.5}			& 26.8			& 86.1			& {\bf 56.3}	& 17.9  \\
& {\tt Human}	    
& 65.4		& 48.9		& 37.3			& 29.1			& {\bf 28.8}			& {\bf 91.3}		& 49.6	& {\bf 21.6} \\
\bottomrule
\end{tabular}
\end{table*}

\subsection{Experimental setup}

\textbf{Dataset and metrics:} We evaluated the proposed method on the AudioCaps dataset \cite{audiocaps}, which consists of audio clips from the AudioSet \cite{audioset} and their captions.
We used the standard split of this dataset; 49838, 495, and 975 audio clips with their captions are provided as training, validation, and testing samples, respectively.
We evaluated the proposed method on the same metrics used in the dataset paper \cite{audiocaps}, i.e., BLEU-1, BLEU-2, BLEU-3, BLEU-4, METEOR, CIDEr, ROUGE-L, and SPICE.

\vspace{5pt}
\noindent
\textbf{Comparison methods:} We compared the proposed method with the systems described in the dataset paper \cite{audiocaps} on three scopes;
(i) the system accuracy,
(ii) the caption retrieval accuracy, and
(iii) the performance upper-bound.

{\bf Scope (i)}: We investigated whether the proposed method enables to generate captions while using a pre-trained language model by comparing it to the state-of-the-art system~\cite{audiocaps}.
As the comparison method, we used {\tt TopDown-AlignedAtt(1NN)} which is the best system proposed in the dataset paper~\cite{audiocaps}.
This system retrieved the nearest training audio from a subset of AudioSet and transfered its labels as conditioning information of the decoder.

{\bf Scope (ii)}: We evaluated the accuracy of guidance caption retrieval. The proposed method is the 1-nearest search with the triplet-based embeddings $\bm{e}$ ({\tt Triplet-Retrieval}), whereas the conventional method is that with raw VGGish embeddings $\bm{\Phi}$ ({\tt 1NN-VGGish}). Both methods find the closest training audio using the $\ell_2$ distance on the embedded features and return its caption as a prediction. As a reference, we also evaluated the scores of the best BERTScore caption as {\tt Top1-BERTScore}.

{\bf Scope (iii)}: We evaluated the performance upper-bound of the proposed method through the accuracy when the estimation accuracy of the audio-based guidance caption retrieval is perfect. We used top-$K$ caption in BERTScores as $\bm{w}^{\mbox{\scriptsize ref}}$. The comparison method was {\tt AlignedAtt(GT)-VGGish-LSTM} which the ground-truth AudioSet labels as the conditioning feature of the decoder. 
As a reference, we also listed the scores of human's captions as {\tt Human} which is the cross validation on the five ground-truth captions~\cite{audiocaps}.

\vspace{5pt}
\noindent
\textbf{Training details (guidance caption retrieval):} 
All trainable parameters were initialized using a random number from $\mathcal{N} (0, 0.02)$ \cite{init}. 
We used dropout before the Transformer-encoder layer with probability $0.3$, and the AdaBound optimizer \cite{adabound} with the initial and final learning rate were 1e-4 and 0.1, respectively.
We train for 200 epochs on minibatches of 128 randomly sampled, and the best validation model was used as the final output.

\vspace{5pt}
\noindent
\textbf{Training details (caption generation):} All captions were tokenized using the word tokenizer of GPT-2, thus the vocabulary size was 50,257. 
All trainable parameters were initialized using a random number from $\mathcal{N} (0, 0.02)$ \cite{init}. 
We used dropout after both multi-head attention layers with probability $0.3$. 
We used the Adam optimizer \cite{adam} with $\beta_1 = 0.9$, $\beta_2 = 0.999$, and $\epsilon = 10^{-8}$ and varied the learning rate was annealed using a cosine schedule where the period was 20 epochs and the maximum and minimum values were 1e-4 and 1e-6, respectively.
We train for 200 epochs on minibatches of 512 randomly sampled, and the best validation model was used as the final output.
We used the beam-search decoding for generating tokens from posteriors where the beamsize was 4.

\vspace{-4pt}
\subsection{Results}

Table \ref{tab:example} shows the evaluation results on the AudioCaps dataset. These results suggest the following:

{\bf Scope (i)}: The proposed method achieved results similar to those of the conventional method which used a carefully designed DNN architecture by the authors of the dataset, even though the proposed method added only a few additional trainable layers after the pre-trained models.
In particular, the proposed method outperformed the conventional method for BLEU-1 and 2 which represent the accuracy of the uni- and bi-grams, and achieved quite close scores for BLEU-3 and METEOR which are that of trigram and the harmonic mean of unigram precision and recall, respectively.
These results suggest that the words in guidance captions can be accurately used to generate captions.
Meanwhile, the proposed method was clearly worse than the conventional method for BLEU-4 and CIDEr which are affected by higher order (longer) $n$-grams concordance rate.
This might be because there is a difference between the higher order $n$-grams of frequently appear in captions and that of the ``free-style'' texts that exist on the web and were used in GPT-2 training.
As a future work, this problem might be solved by fine tuning GPT-2 using training captions.

{\bf Scope (ii)}: The proposed method outperformed the conventional method for all metrics. The difference between two methods is only the embedding network trained using the triplet-loss.
Therefore, it is effective to train an additional embedding network so as to decrease the distance  between the embeddings of an audio pair where its captions' BERTScore is high.

{\bf Scope (iii)}: The proposed method clearly outperformed the conventional method for all metrics. In addition, the proposed method also outperformed the scores of human's caption except three metrics.
This result suggests that adopting a large-scale pre-trained language model to audio captioning has the potential to greatly increase the accuracy of caption generation.

By considering the results of all scopes simultaneously, the block with the greatest promise of improved accuracy is audio-based guidance caption retrieval.
In scope (ii), there is a huge difference in scores between the proposed method and top-1 caption of BERTScore, and there is also a large difference in scores between {\tt Ours} and {\tt Ours (GT)}.
These results suggest that there is a significant room for improvement in the performance of audio-based guidance caption retrieval, and these improvements would improve the performance of caption generation using a large-scale pre-trained language model. Thus, in the future, we will improve the network architecture and the training methods to achieve higher accuracy in caption retrieval.

\vspace{5pt}
We show an example of 
five ground-truth captions, top-5 BERTScore captions, $K=5$ retrieved guidance captions, and the generated caption.
Audio-based guidance caption retrieval successfully searched ``engine'' and ``honk'', however, incorrectly searched ``vehicle'' instead of ``bus'', that results in ``vehicle'' became the subject of the generated caption.

\vspace{2pt}
\noindent
\textbf{Ground truth:}
An engine rumbles loudly, then an air horn honk three times. 
Humming of an engine followed by some honks of a horn.
A bus engine running followed by a bus horn honking.
A bus engine accelerating followed by a bus horn honking while plastic clacks.
A bus engine running followed by a vehicle horn honking.

\vspace{2pt}
\noindent
\textbf{BERTScore top-5:}
A vehicle running followed by a horn honking.
A bus engine running followed by a horn honking and a person laughing.
A muffled bus engine running followed by a vehicle horn honking then a kid laughing.
An engine running followed by a loud horn honking.
A man talking followed by a vehicle horn honking.

\vspace{2pt}
\noindent
\textbf{Guidance captions:}
Traffic flows, brakes squeak, a car horn is honked.
A vehicle engine running as a loud horn honk followed by a softer horn honking.
Vehicle running and honking horn.
Some vehicles move and horn is triggered.
Humming of an engine followed by honking of a car horn.

\vspace{2pt}
\noindent
\textbf{Generated caption:} 
A vehicle engine running followed by a horn honking.

\vspace{0pt}
\section{Conclusions}
\label{sec:cncl}
\vspace{-4pt}

In this study, we examined the use of a pre-trained large-scale language model in audio captioning. In order to overcome the lack of the amount of training data problem, we used GPT-2 as a part of the caption generation, i.e. decoder.
The proposed method consisted of two modules.
The first module retrieved guidance captions for the decoder based on audio similarity. The second module generated a caption using GPT-2 while referring to the retrieved captions.
Experimental results showed that (i) the proposed method had succeeded to use a pre-trained language model for audio captioning, and (ii) the oracle performance of the pre-trained model-based caption generator was clearly superior to the conventional method trained from scratch.
Thus, we conclude that the use of a pre-trained language models is a promising strategy with room for improving the caption generation accuracy.

\clearpage
\bibliographystyle{IEEEbib}
\bibliography{refs}

\begin{thebibliography}{99}
\vspace{-6pt}

\bibitem{ac1} K.\ Drossos, S.\ Adavanne, and T.\ Virtanen,
``Automated Audio Captioning with Recurrent Neural Networks,''
in \textit{Proc. IEEE Workshop Appl. Signal Process. Audio Acoust. (WASPAA)}, 2017.

\bibitem{ac2} S.\ Ikawa and K.\ Kashino,
``Neural Audio Captioning based on Conditional Sequence-to-Sequence Model,''
in \textit{Proc. Detect. Classif. Acoust. Scenes Events (DCASE) Workshop}, 2019.

\bibitem{ac3} M.\ Wu, H.\ Dinkel, and K.\ Yu,
``Audio Caption: Listen and Tell,''
in \textit{Proc. Int. Conf. Acoust. Speech Signal Process. (ICASSP)}, 2019.

\bibitem{audiocaps} C.\ D.\ Kim, B.\ Kim, H.\ Lee, and G.\ Kim,
``AudioCaps: Generating Captions for Audios in The Wild,''
in \textit{Proc. N. Am. Chapter Assoc. Comput. Linguist.: Hum. Lang. Tech. (NAACL-HLT)}, 2019.

\bibitem{clotho} K.\ Drossos, S.\ Lipping, and T.\ Virtanen,
``Clotho: An Audio Captioning Dataset,''
in \textit{Proc. Int. Conf. Acoust. Speech Signal Process. (ICASSP)}, 2020.

\bibitem{ntt_task6} 
Y.\ Koizumi, D.\ Takeuchi, Y.\ Ohishi, N.\ Harada, and K.\ Kashino,
``The NTT DCASE2020 Challenge Task 6 System: Automated Audio Captioning with Keywords and Sentence Length Estimation,''
in \textit{Tech. Rep. Detect. Classif. Acoust, Scenes Events Chall.}, 2020.

\bibitem{Interspeech2020Koizumi}
Y.\ Koizumi,  R.\ Masumura,  K.\ Nishida,  M.\ Yasuda, and S.\ Saito,
``A Transformer-based Audio Captioning Model with Keyword Estimation,''
in \textit{Proc. Interspeech,} 2020.

\bibitem{DCASE2020_Takeuchi}
D.\ Takeuchi, Y.\ Koizumi, Y.\ Ohishi, N.\ Harada, and K.\ Kashino,
``Effects of Word-frequency based Pre- and Post- Processings,''
in \textit{Proc. Detect. Classif. Acoust. Scenes Events (DCASE) Workshop}, 2020.



\bibitem{aed} A.\ Mesaros, T.\ Heittola, A.\ Eronen, and T.\ Virtanen,
``Acoustic Event Detection in Real Life Recordings,''
in \textit{Proc. Euro. Signal Process. Conf. (EUSIPCO)}, 2010.

\bibitem{aed2} K.\ Imoto, N.\ Tonami, Y.\ Koizumi, M.\ Yasuda, R.\ Yamanishi, and Y.\ Yamashita,
``Sound Event Detection By Multitask Learning of Sound Events and Scenes with Soft Scene Labels,''
in \textit{Proc. Int. Conf. Acoust. Speech Signal Process. (ICASSP)}, 2020.

\bibitem{asc} D.\ Barchiesi, D.\ Giannoulis, D.\ Stowell, and M.\ D.\ Plumbley,
``Acoustic Scene Classification: Classifying Environments from the Sounds they Produce,''
\textit{IEEE Signal Process. Mag.}, 2015.

\bibitem{asd} Y.\ Koizumi, S.\ Saito, H.\ Uematsu, Y.\ Kawachi, and N.\ Harada,
``Unsupervised Detection of Anomalous Sound based on Deep Learning and the Neyman-Pearson Lemma,''
\textit{IEEE/ACM Tran. Audio, Speech, and Lang. Process.}, 2019.


\bibitem{vggish} S.\ Hershey, S.\ Chaudhuri, D.\ P.\ W.\ Ellis, J.\ F.\ Gemmeke, A.\ Jansen, R.\ C.\ Moore, M.\ Plakal, DvPlatt, R.\ A.\ Saurous, B.\ Seybold, M.\ Slaney, R.\ Weiss, and K.\ Wilson,
``CNN Architectures for LargeScale Audio Classification,''
in \textit{Proc. Int. Conf. Acoust. Speech Signal Process. (ICASSP)}, 2017.

\bibitem{openl3} J.\ Cramer, H.-H.\ Wu, J.\ Salamon, and J.\ P.\ Bello,
``Look, Listen and Learn More: Design Choices for Deep Audio Embeddings,''
in \textit{Proc. Int. Conf. Acoust. Speech Signal Process. (ICASSP)}, 2019.

\bibitem{coala}
X.\ Favory, K.\ Drossos, T.\ Virtanen, and X.\ Serra
``COALA: Co-Aligned Autoencoders for Learning Semantically Enriched Audio Representations,''
in \textit{Workshop Self-superv. Audio Speech} at \textit{Int. Conf. Mach. Learn. (ICML)}, 2020.

\bibitem{bert} 
J.\ Devlin, M.-W.\ Chang, K.\ Lee, and K.\ Toutanova,
``BERT: Pre-training of Deep Bidirectional Transformers for Language Understanding,''
in \textit{Proc. N. Am. Chapter Assoc. Comput. Linguist.: Hum. Lang. Tech. (NAACL-HLT)}, 2019.

\bibitem{gpt} 
A.\ Radford, K.\ Narasimhan, T.\ Salimans, and I.\ Sutskever,
``Improving Language Understanding with Unsupervised Learning,''
\textit{Tech. rep., OpenAI,} 2018.

\bibitem{gpt2} 
A.\ Radford, J.\ Wu, R.\ Child, D.\ Luan, D.\ Amodei, and I.\ Sutskever. 
``Language models are unsupervised multitask learners,''
\textit{Tech. rep., OpenAI,} 2019.


\bibitem{seq2seq1} I.\ Sutskever, O.\ Vinyals, and Q.\ V.\ Le,
``Sequence to Sequence Learning with Neural Networks,''
in \textit{Proc. Adv. Neural Inf. Process. Syst. (NIPS),} 2014.

\bibitem{seq2seq2} M.\ T.\ Luong, H.\ Pham, and C.\ D.\ Manning
``Effective Approaches to Attention-based Neural Machine Translation,''
in \textit{Proc. Empir. Methods Nat. Lang. Process. (EMNLP),} 2015.

\bibitem{transformer}
A.\ Vaswani, N.\ Shazeer, N.\ Parmar, J.\ Uszkoreit, L.\ Jones, A.\ N.\ Gomez, L.\ Kaiser, and I.\ Polosukhin,
``Attention Is All You Need,''
in \textit{Proc. Adv. Neural Inf. Process. Syst. (NIPS)}, 2017.



\bibitem{BERTScore} T.\ Zhang, V.\ Kishore, F.\ Wu, K.\ Q.\ Weinberger, and Y.\ Artzi,
``BERTScore: Evaluating Text Generation with BERT,''
in \textit{Proc. of Int. Conf. Learn. Representations (ICLR)}, 2020.

\bibitem{triplet} 
J.\ Wang, Y.\ Song, T.\ Leung, C.\ Rosenberg, J.\ Wang, J.\ Philbin, B.\ Chen, and Y.\ Wu.
``Learning Fine-grained Image Similarity with Deep Ranking,''
in \textit{Proc. IEEE Int. Conf. Comput. Vis. Pattern Recognit. (CVPR)}, 2014.

\bibitem{linguistic6}
Y.\ Ohishi, A.\ Kimura, T.\ Kawanishi, K.\ Kashino, D.\ Harwath and J.\ Glass
``Trilingual Semantic Embeddings of Visually Grounded Speech with Self-attention Mechanisms''
in \textit{Proc. Int. Conf. Acoust. Speech Signal Process. (ICASSP)}, 2020.

\bibitem{YasudaInterspeech}
M.\ Yasuda, Y.\ Ohishi, Y.\ Koizumi, and N.\ Harada,
``rossmodal Sound Retrieval based on Specific Target Co-occurrence\\
Denoted with Weak Labels,''
in \textit{Proc. Interspeech,} 2020.

\bibitem{semihard1}
F.\ Schroff, D.\ Kalenichenko, and J.\ Philbin,
``FaceNet: A Unified Embedding for Face Recognition and Clustering'
in \textit{Proc. IEEE Int. Conf. on Comput. Vis. Pattern Recognit. (CVPR)}, 2016.


\bibitem{audioset} J.\ F.\ Gemmeke, D.\ P.\ W.\ Ellis, D.\ Freedman, A.\ Jansen, W.\ Lawrence, R.\ C.\ Moore, M.\ Plakal, and M.\ Ritter,
``Audio Set: An Ontology and Human-Labeled Dataset for Audio Events,''
in \textit{Proc. Int. Conf. Acoust. Speech Signal Process. (ICASSP)}, 2017.

\bibitem{init}
A.\ Radford, K.\ Narasimhan, T.\ Salimans, and I.\ Sutskever,
``Improving Language Understanding by Generative Pre-Training,''
\url{https://blog.openai.com/language-unsupervised}, 2018.

\bibitem{adabound}
L.\ Luo,Y.\ Xiong, Y.\  Liu, and X.\ Sun,
``Adaptive Gradient Methods with Dynamic Bound of Learning Rate,''
in \textit{Proc. Int. Conf. Learn. Representations (ICLR)}, 2019.

\bibitem{adam} D.\ P.\ Kingma and J.\ L.\ Ba,
``Adam: A method for stochastic optimization,''
in \textit{Proc. Int. Conf. Learn. Representations (ICLR)}, 2015.


\end{thebibliography}

\end{document}